\documentclass[aps,prb,twocolumn,floatfix]{revtex4}%
\bibpunct{[}{]}{,}{n}{}{}

\usepackage{graphicx}
\usepackage{dcolumn}
\usepackage{latexsym}
\usepackage{amssymb}
\usepackage{amsfonts}
\usepackage{amsmath}
\usepackage{fancybox}
\usepackage[binary,squaren]{SIunits}
\usepackage{color}
\usepackage{colordvi}%
\providecommand{\U}[1]{\protect\rule{.1in}{.1in}}

\def\be{\begin{equation}}
\def\ee{\end{equation}}
\def\ber{\begin{eqnarray}}
\def\eer{\end{eqnarray}}

\allowdisplaybreaks[2]
\begin{document}
\title{Finite temperature magnon spectra in yttrium iron garnet from mean field approach in tight-binding model}

\author{Ka Shen}
\affiliation{The Center for Advanced Quantum Studies and Department of Physics, Beijing Normal University, Beijing 100875, China}


\date{\today}

\begin{abstract}
We study magnon spectra at finite temperature in yttrium iron garnet from tight-binding model with nearest neighboring exchange interaction. The spin reduction due to thermal magnon excitation are taken into account via the mean field approximation to the local spin and found to be different at two sets of iron atoms. The resulting temperature dependence of the spin wave gap shows good agreement with experiment. We find only two magnon modes are relevant to ferromagnetic resonance.

\end{abstract}


\maketitle

\section{Introduction}
Since its discovery decades ago~\cite{Geller57}, yttrium iron garnet (YIG) has been regarded as one of the most important magnetic materials due to its extremely low magnetic damping and other intriguing properties~\cite{Cherepanov93}. As an insulator, YIG is free from Joule heating, revealing its promising applications in future low energy consumption devices, which makes it a popular material for recent study on magnonics~\cite{Serga10,Chumak15} and spin caloritronics~\cite{Bauer12}. While the applications in magnonics mainly rely on the coherent transport properties of spin waves with relatively long wavelength ($\sim\mu$m) and low frequency ($\sim$GHz), in spin caloritronic devices, the short wavelength spin waves, namely, magnons, dominate because of their significant population from thermal excitation. In the latter case, recent experiments show that the diffusion length of thermal magnons can reach tens of microns~\cite{Giles15,Cornelissen15} and present very interesting behaviors with respect to variation of external magnetic field and temperature~\cite{Cornelissen16,Cornelissen16b,Cornelissen16c} which have not been fully understood so far. Moreover, the magnon diffusion length is also found to be sensitive to the method of measurement~\cite{Giles15,Cornelissen15}. The full understanding on those observation obviously requires a comprehensive study of dissipation process beyond the long wavelength limit into a wide range of magnon frequency up to THz.

In the literature, great efforts~\cite{Schlomann61,Sparks64,Akhiezer68} have been devoted to explain the origin of magnetic damping in YIG measured by ferromagnetic resonant~\cite{Spencer59,Spencer60,Spencer60b,Nemarich64,Roschmann75,Klingler17} or parametric pumping~\cite{Kasuya61,Gurevich75}. Several mechanisms are believed to be relevant, such as magnon-impurity scattering, magnon-phonon scattering, and magnon-magnon interaction as well as spin-lattice relaxation. However, both the uniform mode in ferromagnetic resonant experiment and finite wave vector magnon from parametric pumping lie in GHz regime, therefore, the applicability of the conclusions from GHz magnons to thermal magnons of THz is questionable. In some recent works, the lifetime of the thermal magnons in YIG were estimated with a single parabolic dispersion~\cite{Cornelissen16}, which however is valid only for GHz magnons and could depart far from the real spectra. Moreover, at room temperature, several magnon bands could be excited in YIG~\cite{Plant77,Barker16}, which again reveals the limitation of the single band model.

The full magnon band structure of YIG were modeled by Harris~\cite{Harris63} within nearest neighboring exchange interaction and measured from neutron scattering by Plant~\cite{Plant77}, from which the exchange interaction coefficients were fitted out~\cite{Cherepanov93}. The next nearest neighboring exchanges were recently also taken into account with the exchange integrals calculated from first principle calculation~\cite{Xie17} or fitted to neutron scattering data~\cite{Plant83,Princep17}. Atomistic dynamic simulation shows that the nearest neighboring exchange model with stochastic thermal fluctuation is sufficient to quantitatively reproduce the decrease of the spin wave gap, minimal frequency of the antiferromagnetic-like mode, with increasing temperature~\cite{Barker16}. The reason for such a red shift behavior is that the effective magnetic field a local spin felt is suppressed by the reduction of the expectation value of neighboring spins due to magnon excitation. In the low temperature regime, the reduction of average spin is too small to be relevant, resulting in saturate spin wave gap and magnetization, observed in experiment. The gap and magnetization from the atomistic dynamic simulation of classical spin however keep linearly increasing and fails to saturate at very low temperature~\cite{Barker16}, in contrast to the results from experiment and Harris's quantum model with the same parameters. 

In the present work, we develop a mean field approach based on nearest neighboring exchange model, where the magnon excitation at finite temperature is taken into account. In our approach, the temperature dependence of the entire magnon spectra can be easily obtained by selfconsistently solving the magnon population and its influence on dispersion. With temperature independent exchange constants, the resulting spin wave gap and magnetization show good agreement with experiment from zero up to room temperature. In addition to the proper description of the entire spectra beyond the widely used long wavelength approximation, another apparent advantage of our approach is that the output tight-binding type wave functions contain only twenty components, which is therefore convenient for quantitative computation of various spin wave properties, e.g., spin wave dissipation due to different mechanisms, as well as the responses to different sources, e.g., light or microwave. Moreover, the information of the polarization, responsible to spin Seebeck effect~\cite{Barker16}, is also naturally included in the wave functions and easy to read out. Therefore our approach should be useful for study various physics in YIG at finite temperature. As an application, we discuss ferromagnetic resonance and find only two of twenty modes are relevant. Our approach can also be extended to nano structure.

\section{Exchange model}
In one unit cell in YIG, the collinearly aligning iron atoms  at ground state are separated into two sets, eight $a$ atoms and twelve $d$ atoms, according to the polarization or the configuration of neighboring oxygen atoms~\cite{Cherepanov93}. The closest distances are $r_{aa}=(\sqrt{3}/4)a_0$ between two $a$-site atoms, $r_{dd}=(\sqrt{6}/8)a_0$ between two $d$-atoms, and $r_{ad}=(\sqrt{5}/8)a_0$ between $a$ and $d$ atoms with $a_0$ being the lattice constant of face-centered cubic lattice. Following Harris~\cite{Harris63}, we start from the nearest neighboring exchange Hamiltonian
\ber
H&=&-\sum_{n=1}^N\Big[J_{aa}\sum_{i}^{8}\sum_{|\mathbf r_{ij}|=r_{aa}}\mathbf{S}_{a}(\mathbf{R}_{in})\cdot\mathbf{S}_{a}(\mathbf{R}_{in}+\mathbf{r}_{ij})\nonumber \\
&&+2J_{ad}\sum_{i=1}^{8}\sum_{|\mathbf r_{ij}|=r_{ad}}\mathbf{S}_{a}(\mathbf{R}_{in})\cdot\mathbf{S}_{d}(\mathbf{R}_{in}+\mathbf{r}_{ij})\nonumber\\
&&+J_{dd}\sum_{i=9}^{20}\sum_{|\mathbf r_{ij}|=r_{dd}}\mathbf{S}_{d}(\mathbf{R}_{in})\cdot\mathbf{S}_{d}(\mathbf{R}_{in}+\mathbf{r}_{ij})\nonumber\\
&&+\sum_{i=1}^{8}g\mu_B\mathbf{B}\cdot\mathbf{S}_{a}(\mathbf{R}_{in})+\sum_{i=9}^{20}g\mu_B\mathbf{B}\cdot\mathbf{S}_{d}(\mathbf{R}_{in})\Big], \label{Hex}
\eer
where $J_{aa}$, $J_{dd}$, and $J_{ad}$ represent the corresponding exchange coupling constants. $N$ is the number of unit cell. The last two terms are Zeeman energy. In YIG, all these coupling constants are negative~\cite{Cherepanov93}. The coordinates of atoms can be found in Ref.~\cite{Harris63}.

Without loss of generality, we assume that the equilibrium magnetization follows the small external magnetic field along $z$ axis. By defining transverse components $S^{\pm}=S^{x}\pm iS^{y}$, the Holstein-Primakoff transformation of  two antiferromagnetically coupled sublattices can be written as~\cite{Holstein40,Takahashi89}
\ber
S_{a}^{z}=S_{a}-a^{\dagger}a,& S_{a}^{+}=\sqrt{2S_{a}-a^{\dagger}a}a,\nonumber\\
S_{d}^{z}=-S_{d}+d^{\dagger}d, &S_{d}^{+}=d^{\dagger}\sqrt{2S_{d}-d^{\dagger}d},\label{HP}
\eer
and $S_{a,d}^{-}= (S_{a,d}^{+})^\dag$ where we have taken into account the antiferromagnetic alignment of two sublattices. Thus, $a^\dag$ and $d^\dag$ are the creation operators of clockwise rotation in $a$-site and anticlockwise rotation in $d$-site, respectively. 
Applying Eq.~(\ref{HP}) into Eq.~(\ref{Hex}) and Fourier transform $a_{in}=\frac{1}{\sqrt{N}}\sum_{\mathbf{k}}a_{i}(\mathbf{k})e^{i\mathbf{k}\cdot\mathbf{R}_{in}}$, one obtains the tight-binding type Bogoliubov-de Gennes Hamiltonian for an arbitrary wave vector $\mathbf k$ (the quadratic order)~\cite{Harris63,Cherepanov93}
\ber
H_{\mathbf k}&=&\sum_{i,j=1}^{8}A_{ij}(\mathbf{k})a_{i}^{\dagger}(\mathbf{k})a_{j}(\mathbf{k})+\sum_{i,j=9}^{20}D_{ij}(\mathbf{k})d_{i}^{\dagger}(\mathbf{k})d_{j}(\mathbf{k})\nonumber\\
&&+\sum_{i=1}^{8}\sum_{j=9}^{20}[B_{ij}(\mathbf{k})a_{i}^{\dagger}(\mathbf{k})d_{j}^{\dagger}(-\mathbf{k})+h.c.], \label{Hk}
\eer
where
\ber
A_{ij}(\mathbf{k})&=&(-12J_{ad}S_{d}+16J_{aa}S_{a}+g\mu_BB)\delta_{ij}\nonumber\\
&&-2J_{aa}S_{a}\gamma_{ij}^{aa}(\mathbf{k}),\nonumber \\
D_{ij}(\mathbf{k})&=&(-8J_{ad}S_{a}+8J_{dd}S_{d}-g\mu_BB)\delta_{ij}\nonumber\\
&&-2J_{dd}S_{d}\gamma_{ij}^{dd}(\mathbf{k}),\nonumber\\
B_{ij}(\mathbf{k})&=&-2J_{ad}\sqrt{S_{a}S_{d}}\gamma_{ij}^{ad}(\mathbf{k}),
\eer
with $\gamma_{ij}^{\eta\eta'}(\mathbf{k})=\sum_{|\boldsymbol{r}_{ij}|=r_{\eta\eta'}}e^{i\mathbf{k}\cdot\mathbf{r}_{ij}}$. The last term in Eq.~(\ref{Hk}) leads to angular momentum exchange between the two sublattices preserving the chirality.

\section{Results}
In this section, we present our results from the diagonalization of the above Bogoliubov-de Gennes Hamiltonian. The exchange constants are specified to be $J_{ad}=-39.8$~K, $J_{dd}=-13.4$~K, and $J_{aa}=-3.8$~K~\cite{Cherepanov93}. Both $a$- and $d$-sites are approximately of $S_a=S_d=5/2$. Note that, instead of unitary transform in fermionic system, a paraunitary transform is needed in the diagonalization procedure of bosonic Hamiltonian matrix under basis $(a_{\mathbf k},d_{\mathbf k},a^\dag_{-\mathbf k},d^\dag_{-\mathbf k})$~\cite{Colpa78,Benedetta_polaron}. The resulting eigenstates contains two sets
\ber
\alpha_{i\boldsymbol{k}}&=&c_{1,\mathbf k}^{ii'}a_{i'\mathbf k}+c_{2,\mathbf k}^{ij'}d_{j'-\mathbf k}^{\dagger},\\
\beta_{j\boldsymbol{k}}&=&c_{3,\mathbf k}^{ji'}a_{i'-\mathbf k}^{\dagger}+c_{4,\mathbf k}^{jj'}d_{j'\mathbf k},
\eer
which satisfy the Boson commutation rule 
\ber
&[\alpha_{i\mathbf k},\alpha^\dag_{i'\mathbf k'}]=\delta_{ii'}\delta_{\mathbf k\mathbf k'},\\
&[\beta_{j\mathbf k},\beta^\dag_{j'\mathbf k'}]=\delta_{jj'}\delta_{\mathbf k\mathbf k'},
\eer
guaranteed by the paraunitary transform. Einstein summation convention has been applied with $i,i'=1,...,8$ and $j,j'=1,...,12$. Since the creation of an anticlockwise rotation of $d$-site ($d^\dag$) is equivalent to the annihilation of a clockwise rotation, $\alpha_{i\mathbf k}$ are annihilation operators of pure clockwise rotation. In the same sense, $\beta_{j\mathbf k}$ are about purely anticlockwise rotation. The dynamics of the transverse components of local spin for two types of modes read
\ber
(S_{i'a}^{x},S_{i'a}^{y})_{\alpha_{i\boldsymbol{k}}}(t)&=&(\Re c_{1,\mathbf{k}}^{ii'}(t),\Im c_{1,\mathbf{k}}^{ii'}(t))\\
(S_{j'd}^{x},S_{j'd}^{y})_{\alpha_{i\boldsymbol{k}}}(t)&=&(\Re c_{2,\mathbf{k}}^{ij'}(t),\Im c_{2,\mathbf{k}}^{ij'}(t))\\
(S_{i'a}^{x},S_{i'a}^{y})_{\beta_{j\boldsymbol{k}}}(t)&=&(\Re c_{3,\mathbf{k}}^{ji'}(t),-\Im c_{3,\mathbf{k}}^{ji'}(t))\\
(S_{j'd}^{x},S_{j'd}^{y})_{\beta_{j\boldsymbol{k}}}(t)&=&(\Re c_{4,\mathbf{k}}^{ij'}(t),-\Im c_{4,\mathbf{k}}^{ij'}(t))
\eer
where the time evolution of the coefficients are
$c_{1,\mathbf{k}}^{ii'}(t)=e^{i\boldsymbol{k}\cdot\boldsymbol{r}_{i'}-i\omega_{\alpha_{i\boldsymbol{k}}}t}c_{1,\mathbf{k}}^{ii'}$,
$c_{2,\mathbf{k}}^{ij'}(t)=e^{-i\boldsymbol{k}\cdot\boldsymbol{r}_{j'}-i\omega_{\alpha_{i\boldsymbol{k}}}t}c_{2,\mathbf{k}}^{ij'}$,
$c_{3,\mathbf{k}}^{ji'}(t)=e^{-i\boldsymbol{k}\cdot\boldsymbol{r}_{i'}-i\omega_{\beta_{j\boldsymbol{k}}}t}c_{3,\mathbf{k}}^{ji'}$,
$c_{4,\mathbf{k}}^{ij'}(t)=e^{i\boldsymbol{k}\cdot\boldsymbol{r}_{j'}-i\omega_{\beta_{j\boldsymbol{k}}}t}c_{4,\mathbf{k}}^{ij'}$.
The inverse transform gives the original site-operators in form of a combination of magnon operators of eigenstates
\ber
a_{i'\boldsymbol{k}}&=&C_{1,\mathbf k}^{i'i}\alpha_{i\mathbf k}+C_{2,-\mathbf k}^{i'j}\beta_{j-\mathbf k}^{\dagger},\\
d_{j'\boldsymbol{k}}&=&C_{3,-\mathbf k}^{j'i}\alpha_{i-\mathbf k}^{\dagger}+C_{4,\mathbf k}^{j'j}\beta_{j\mathbf k}.
\eer
The fluctuation of local spins is then given by 
\ber
\langle  a_{i'}^\dag a_{i'}\rangle &=&\frac{1}{N}\sum_{\mathbf k}|C_{1,\mathbf k}^{i'i}|^2 \langle \alpha_{i\mathbf k}^\dag \alpha_{i\mathbf k}\rangle + |C_{2,\mathbf k}^{i'j}|^2 (\langle \beta_{j\mathbf k}^\dag \beta_{j\mathbf k}\rangle+1),\nonumber\\
\\
\langle  d_{j'}^\dag d_{j'}\rangle &=&\frac{1}{N}\sum_{\mathbf k}|C_{3,\mathbf k}^{j'i}|^2 (\langle \alpha_{i\mathbf k}^\dag \alpha_{i\mathbf k}\rangle +1)
+ |C_{4,\mathbf k}^{j'j}|^2 \langle \beta_{j\mathbf k}^\dag \beta_{j\mathbf k}\rangle,\nonumber\\
\eer
which shows that at very low temperature, where the thermal excitation is negligible, i.e., $\langle \alpha_{i\mathbf k}^\dag \alpha_{i\mathbf k}\rangle\approx\langle \beta_{j\mathbf k}^\dag \beta_{j\mathbf k}\rangle\approx0$, there is still non-zero fluctuation due to the coupling between $a$- and $d$-site rotations via $B_{ij}$ terms in Eq.~(\ref{Hk})~\cite{Akashdeep17,Akashdeep17B}.

\begin{figure}[]
  \includegraphics[width=4.cm]{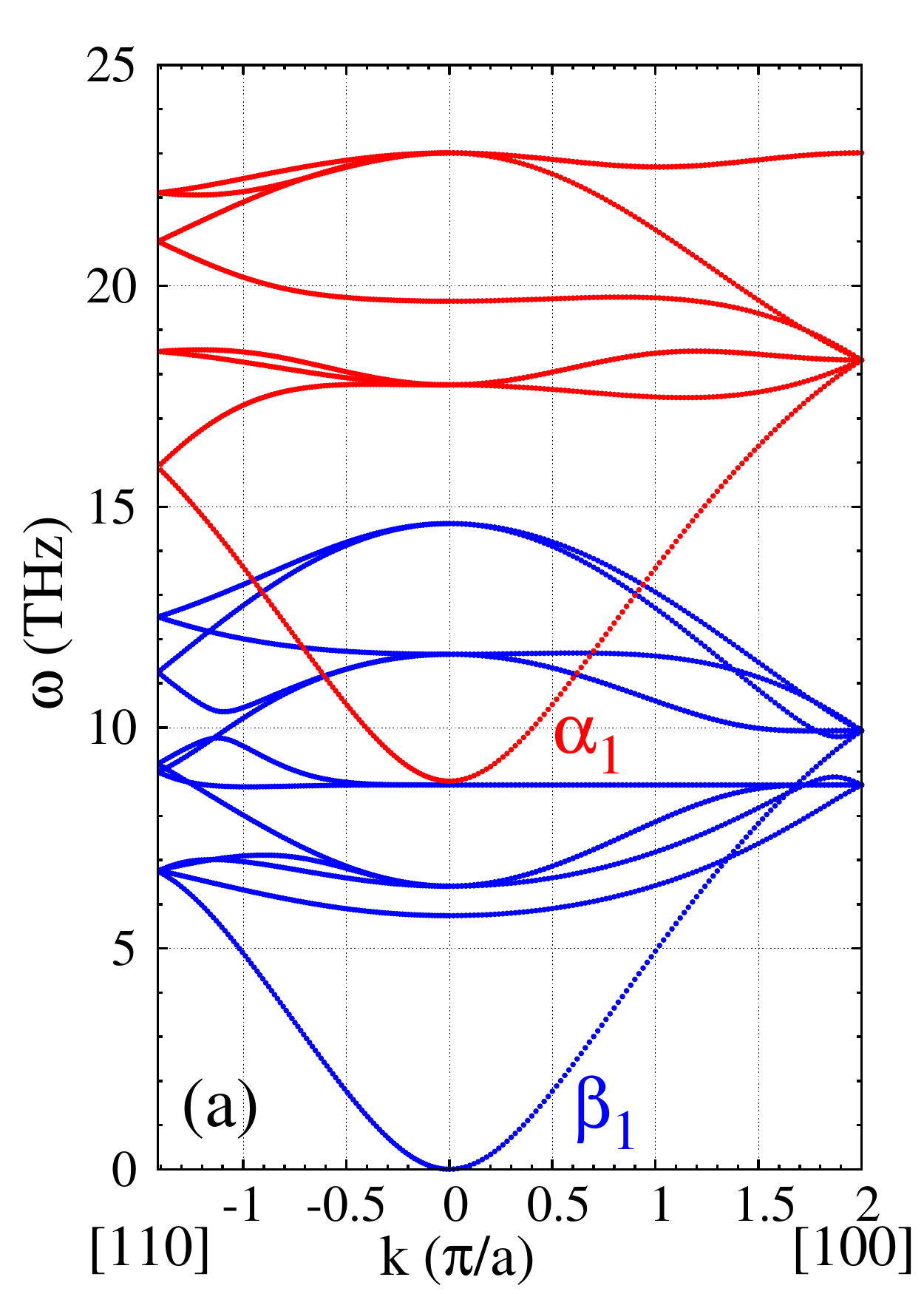}
  \includegraphics[width=4.cm]{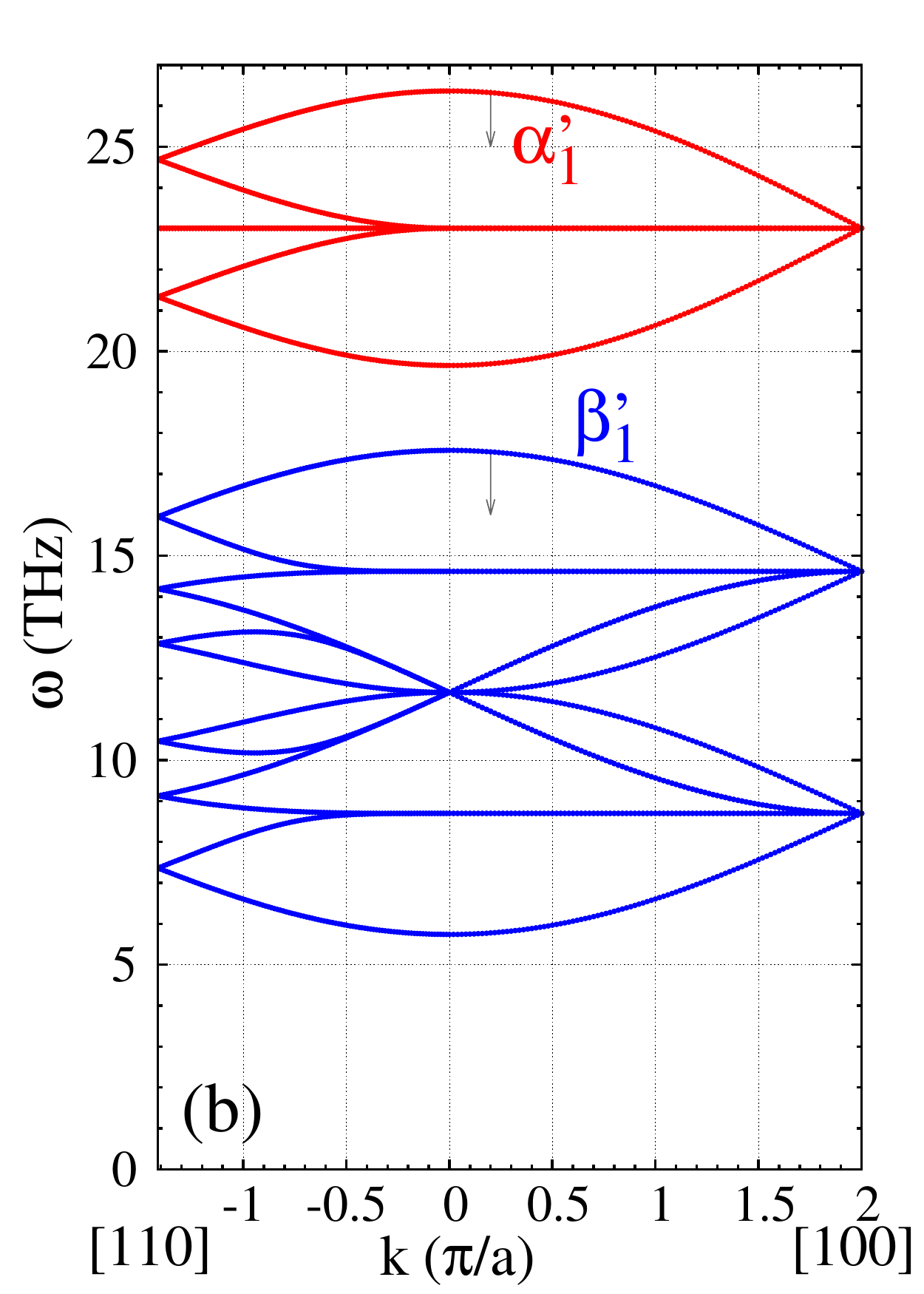}
  \includegraphics[width=8.6cm]{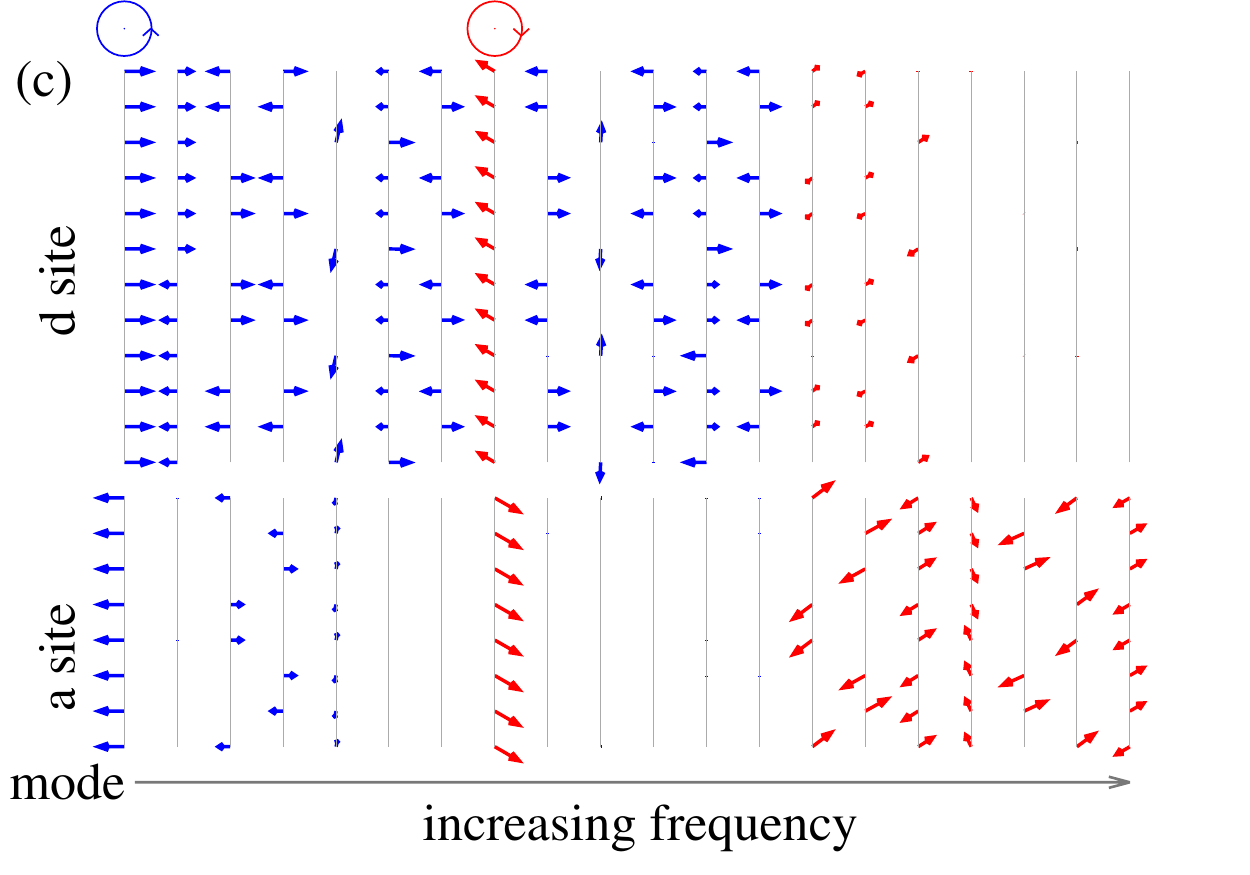}
  \caption{Zero temperature spin wave spectra (a) with and (b) without inter-sublattice coupling $B_{ij}$. The blue and red colors separate spin waves into two group according to polarizations. (c) Instantaneous spin configuration of all twenty modes in (a) near $k=(0,0,0)$.}
\label{T0}
\end{figure}

\subsection{Zero temperature spectra}
Fig.~\ref{T0}(a) shows all twenty magnon branches in [110] and [100] directions carried out from zero temperature in the absence of external magnetic field. The characteristics of polarization $\alpha_{i\mathbf k}$ and $\beta_{i\mathbf k}$ are distinct from the color red and blue, respectively. Typically, the eight clockwise modes have higher energy than the twelve anticlockwise modes, since the former correspond to the $a$-site spins rotating under the exchange field of six nearby $d$-site spins, which is stronger ($\sim 12 |J_{ad}|S_d$) than the one felt by the $d$-site spin from the four nearby $a$-site spins ($\sim 8|J_{ad}|S_a$). This is confirmed by Fig.~\ref{T0}(b) from the calculation with the same parameters but setting all $B_{ij}=0$, therefore, no coupling between clockwise and anticlockwise rotations. From comparison between the results with and without $B_{ij}$ terms, we can see that the angular momentum exchange between the sublattices partially lifts the degeneracy at high symmetry points, and more importantly leads to the correct dispersion of acoustic mode and the antiferromagnetic-like mode, labeled as  $\beta_1$ and $\alpha_1$, separately. The $\beta_1$ mode is almost isotropic up to 5~THz and starts to become anisotropic approaching the boundary of the Brillouin zone.

In Fig.~\ref{T0}(c) we plot the instantaneous spin configuration of each mode at the center of Brillouin zone ($\mathbf k=0$). The transverse spins in both $\alpha_1$ and $\beta_1$ are parallel within the sublattices and antiparallel between sublattices. The magnitudes of spins in $\beta_1$ mode are all the same, as a result, this tilted configuration is a ground state of the system and stable, explaining its zero frequency. The $a$ spins in $\alpha_1$ contain larger transverse spin components and therefore dominate the rotation direction of this mode. Without $B_{ij}$, the two sublattices are decoupled, as a result, the parallel configuration of each sublattice ($\alpha_1'$ and $\beta_1'$) is of highest energy among the separately clockwise and anticlockwise modes due to negative exchange coupling constant between the same type of atoms [see Fig.~\ref{T0}(b)]. 

\begin{figure*}[]
\centering
  \makebox[\textwidth][c] {
  \includegraphics[width=5cm]{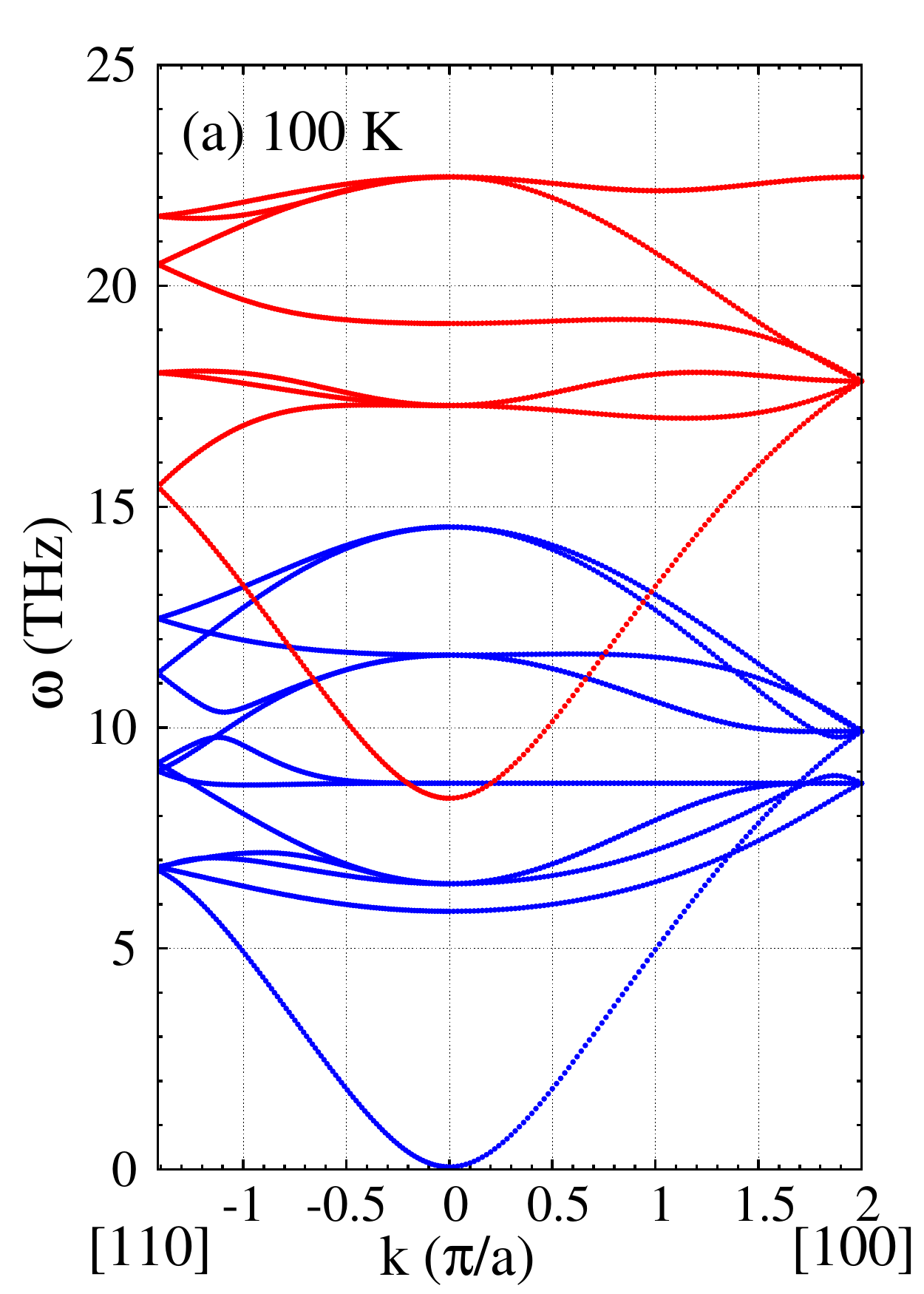}
  \includegraphics[width=5cm]{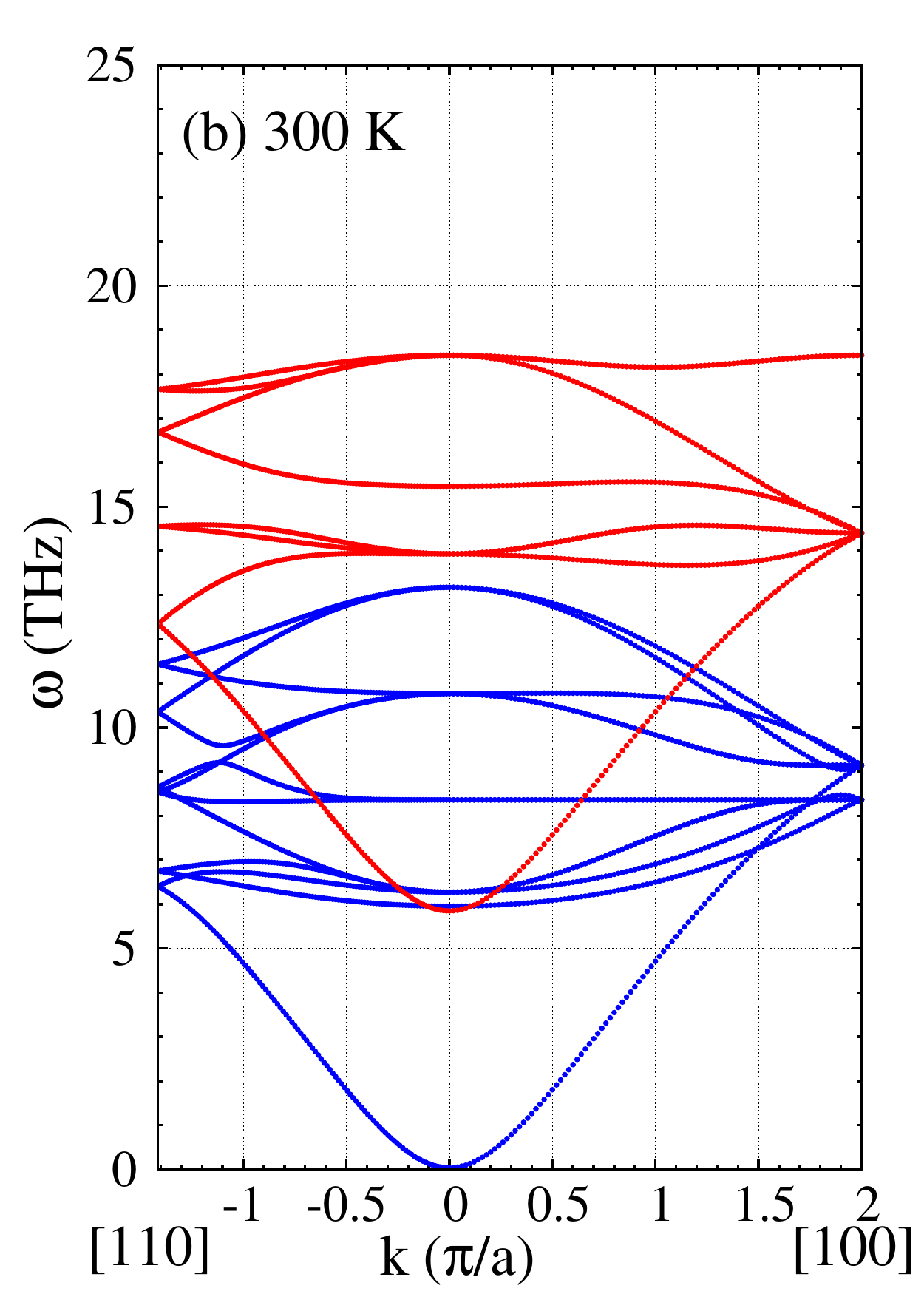}
  \includegraphics[width=5.2cm]{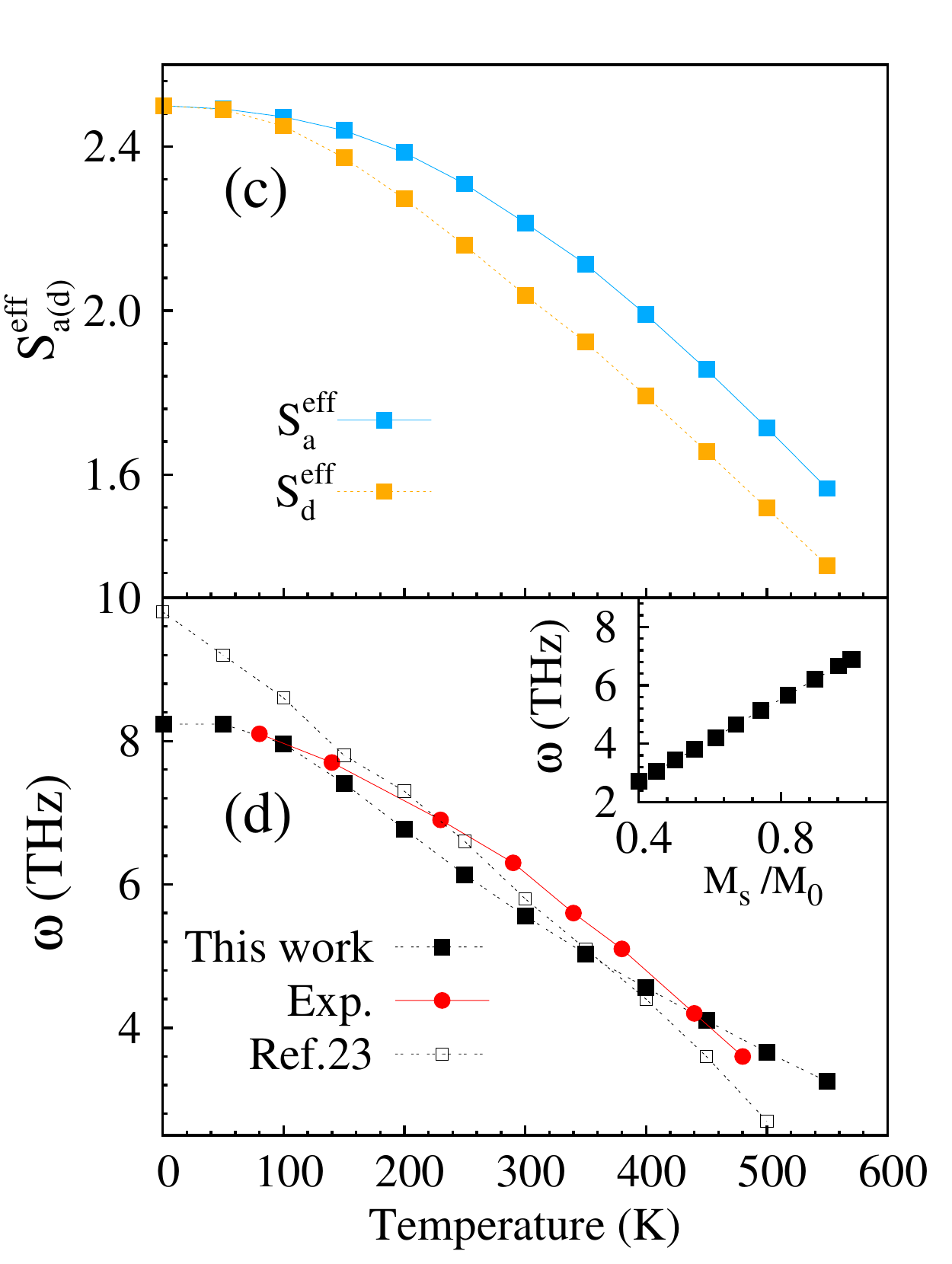}
}
  \caption{Spin wave spectra at (a) 100~K and (b) 300~K. (c) Effective local spin at $a$ and $d$ as function of temperature. (d) Minimal frequency of antiferromagnetic-like mode as function of temperature. The inset gives its dependence on magnetization.}
\label{Tdep}
\end{figure*}

\subsection{Finite temperature spectra}
At low temperature, the density of the thermally excited magnon is low enough so that the magnon dispersion calculated at zero temperature still works well. However as more magnons are excited at high temperatures, the enhanced magnon-magnon interaction can strongly affect the magnon dispersion. Physically, for example, the large magnon density reduces the magnitude of the expectation value of all local spins, hence the exchange field felt by the neighboring spins. The effect of magnon-magnon interaction can be taken into account by including the higher order terms beyond quadratic approximation, e.g., those in form of $a_i^\dag a_i^\dag a_i a_j$,  $a_i^\dag a^\dag_i a_i d_j$, $a_i^\dag a_i d^\dag_j d_j$,  etc. The modification in magnon spectra then can be carried out by diagrammatic calculation on the real part of the magnon self-energy due to these interaction terms. Here, alternatively, we perform a mean field approximation with a self-consistent procedure, whose advantage has been discussed in the Introduction.

Specifically, we keep all magnon-magnon interactions with up to three magnon operators at the same site and apply
 \ber
a^\dag_i a_i\to \langle  a^\dag_i a_i\rangle,&&
a^\dag_i a_i^\dag a_i\to 2\langle  a^\dag_i a_i \rangle a^\dag_i\\
d^\dag_j d_j\to \langle  d^\dag_j d_j\rangle ,&&
d^\dag_j d_j^\dag d_j\to 2\langle  d^\dag_j d_j \rangle d^\dag_j
\eer
for interactions between $a$-site spins to reduce the interaction terms back to quadratic order. The resulting effective quadratic Hamiltonian is exactly of the same expression as Eq.~(\ref{Hk}) after a replacement of $S_{a(d)}$ by
\be
S_{a(d)}^{\rm eff}=S_{a(d)}-\overline{\Delta S_{a(d)}}. \label{effS}
\ee
where $\overline{\Delta S_{a(d)}}$ stands for the average of thermal excitation over all $a(d)$ site spins
\ber
\Delta S_{i'a}&=&\frac{1}{N}\sum_{\mathbf k}|C_{1,\mathbf k}^{i'i}|^2n_B(\hbar\omega_{\alpha_{i\mathbf k}})+|C_{2,\mathbf k}^{i'j}|^2n_B(\hbar\omega_{\beta_{i-\mathbf k}}),\nonumber\\
\Delta S_{j'd}&=&\frac{1}{N}\sum_{\mathbf k}|C_{3,\mathbf k}^{j'i}|^2n_B(\hbar\omega_{\alpha_{i-\mathbf k}})+|C_{4,\mathbf k}^{j'j}|^2n_B(\hbar\omega_{\beta_{i\mathbf k}}).\nonumber\\
\eer
Here, $n_B$ corresponds to Plank distribution at thermal equilibrium. By diagonalizing Eq.~(\ref{Hk}) with $S^{\rm eff}_{a(d)}$, we obtain modified spin wave spectra, from which we compute updated $S^{\rm eff}_{a(d)}$ and repeat such process until reaching a self-consistent solution.

In Figs.~\ref{Tdep}(a) and (b), we plot the convergent magnon spectra at 100~K and 300~K, respectively. As we can see, the shape of each band roughly remain the same at different temperatures. However, as the temperature increases, the clockwise rotating mode (in red) move towards to lower frequency regime, which is consistent with experimental observation~\cite{Plant77} and atomistic simulation~\cite{Barker16}. The anticlockwise rotating modes are relatively insensitive to the temperature change.

In Fig.~\ref{Tdep}(c), the effective magnitudes of two sets spins, i.e., Eq.~(\ref{effS}), are plotted as function of temperature, which shows weak dependence in the low temperature regime, and an approximately linear decrease in the temperature regime. Interestingly, the reduction at $d$-site is much larger than that at $a$-site. The reason for such behavior is that the $d$ site has larger components than $a$ site in the low energy magnon bands, who are efficiently excited. At room temperature, the effective spin reduces to 2.21 and 2.03 respectively. These distinct reductions are actually the origin of the relative shift between anticlockwise and clockwise sets in Figs.~\ref{Tdep}(a) and (b), because the frequencies of the clockwise modes rely on the magnitude of $d$ spins. If the two effective magnitudes are equal $S^{\rm eff}_{a}=S^{\rm eff}_{d}$, all frequencies would vary simultaneously by a global prefactor according to Eq.~(\ref{Hk}), which is proportional to $S_a^{\rm eff}=S_d^{\rm eff}$ at vanishing magnetic field.

We summarize in Fig.~\ref{Tdep}(d) the frequency at the button of the $\alpha_1$ mode (i.e., the spin wave gap mentioned in the Introduction) at different temperatures as black solid squares. For comparison, the experimental data~\cite{Plant77} and results from atomistic simulation~\cite{Barker16} are plotted in the same figure as red bullet and open squares, respectively. As we can see that our results show good agreement with experiment especially at low temperature, where limited magnon are excited hence approaching to the zero temperature case. Our approach therefore avoids the low temperature deviation in atomistic simulation based on classical spin. In the inset, we use the magnetization from our calculation at finite temperature and plot the frequency as function of the corresponding magnetization of the same temperature, which presents good linear dependence. The magnetization at room temperature is around $\mu_0M_s(300~K)=0.163~T$, being reasonable agreement with experimetal data $0.175~T$~\cite{Manuilov09}. The deviation from experimental value is because of the fact that the temperature dependence of the  magnon spectra was not taken into account in the procedure to determine the exchange constants from temperature dependence of magnetization~\cite{Cherepanov93}. The determination of the correction in exchange constants due to the change in magnon spectra is beyond the scope of the present work.

\section{Ferromagnetic resonance}
As discussed in the Introduction, with the knowledge of wave functions, one can calculate various properties. Here, we take ferromagnetic resonance for an example. Assume spatially uniform ac magnetic field transverse to the magnetization, we write out the time dependent perturbation Hamiltonian to describe its coupling with magnetic atoms
\be
H'(t)=g\mu_B \mathbf B(t)\cdot [\sum_{i,n} \mathbf S_{a}(\mathbf R_{in})+\sum_{j,n} \mathbf S_{d}(\mathbf R_{jn})].
\ee
For linear polarization magnetic field along $x$-direction, 
\be
H'(t)=\sqrt{N}g\mu_BB_{x}(t)({\cal G}^\alpha_{i}\alpha_{i\boldsymbol{0}}+{\cal G}^\beta_{j}\beta_{j\boldsymbol{0}})+h.c.
\ee
with ${\cal G}_{i}^{\alpha}=\sum_{i'}C_{1,\boldsymbol{0}}^{i'i}+\sum_{j'}C_{3,\boldsymbol{0}}^{j'i\ast}$, ${\cal G}_{j}^{\beta}=\sum_{i'}C_{2,\boldsymbol{0}}^{i'j\ast}+\sum_{j'}C_{4,\boldsymbol{0}}^{j'j}$, and $B_{x}(t)=B_0\cos(\omega t)$. Here, only $\mathbf k=0$ survives in the space integration. The absorption rates are then given by Fermi Golden rule
\ber
\Gamma(\omega)&=&\frac{\pi}{2\hbar}(g\mu_BB_0)^2N \big\{|{\cal G}_{i}^{\alpha}|^2[N_B(\omega_{\alpha_{i\boldsymbol{0}}})+1]\delta(\omega-\omega_{\alpha_{i\boldsymbol{0}}})\nonumber\\
&&\hspace{0.5cm}+|{\cal G}_{j}^{\beta}|^2[N_B(\omega_{\beta_{j\boldsymbol{0}}})+1]\delta(\omega-\omega_{\beta_{j\boldsymbol{0}}})\big\},
\eer
which is proportional to the number of unit cells in magnetic sample and the input power ($\propto B_0^2$). By using the coefficients in wave function $C_{1,2,3,4}$, we find only two of the prefactors ${\cal G}_1^\alpha$ and ${\cal G}_1^\beta$ are non-zero. This actually can be easily understood from the spin configuration in Fig.~\ref{T0}(c), where only $\alpha_1$ and $\beta_1$ modes have net spin. The vanishing of net spin in other modes makes them transparent to driving field.

\section{Conclusion and discussion}

In conclusion, we calculated the entire magnon spectra at finite temperature within mean field approximation on the magnon-magnon interaction. The temperature dependence of magnon spectra shows good agreement with experiment. We find the reduction in magnitude of $d$ spin is larger than that of $a$ atoms. By using the wave functions of eigenstates, we analyze the ferromagnetic resonance and find that only two modes can be drived by spatially uniform ac magnetic field. We suppose our approach which provides proper magnon spectra and corresponding wave functions can be usefully to study various properties in yttrium iron garnet at finite temperature as discussed in the Introduction.

One may notice that in Fig.~\ref{Tdep} the residual magnetization (effective local spin) near Curie temperature (560~K) is still finite, which reveals the limitation of the spin wave approximation in the high temperature regime~\cite{Cherepanov93}. In the literature, the mean field theory based on sublattices' magnetizations was found to perform well up to Curie temperature~\cite{Mitchell87}. How to link this theory to our tight-binding type mean field approach and how to extend the present spin-wave-based model up to Curie temperature in ferrimagnetic systems are still under study.

\begin{acknowledgements}
The author thanks G. E. W. Bauer and J. L. Cheng for valuable discussions.
This work is supported by the Recruitment Program of Global Youth Experts. 

\end{acknowledgements}

\bibliographystyle{prsty}

\bibliography{Refs.bib}

\end{document}